\documentclass[11pt,a4paper]{article}

\usepackage{amsmath,amssymb,amsthm}
\usepackage{geometry}
\usepackage{booktabs}
\usepackage{xcolor}
\usepackage{tcolorbox}
\usepackage{microtype}
\usepackage{nicefrac}
\usepackage{bm}
\usepackage{hyperref}

\tcbuselibrary{skins,breakable}

\newtcolorbox{keybox}[1][Key Insight]{
  enhanced, breakable,
  colback=blue!4!white,
  colframe=blue!55!black,
  fonttitle=\bfseries\small,
  title={#1},
  left=6pt, right=6pt, top=4pt, bottom=4pt
}

\newtcolorbox{warnbox}[1][Warning]{
  enhanced, breakable,
  colback=red!4!white,
  colframe=red!55!black,
  fonttitle=\bfseries\small,
  title={#1},
  left=6pt, right=6pt, top=4pt, bottom=4pt
}

\newtcolorbox{resultbox}[1][Result]{
  enhanced, breakable,
  colback=green!4!white,
  colframe=green!50!black,
  fonttitle=\bfseries\small,
  title={#1},
  left=6pt, right=6pt, top=4pt, bottom=4pt
}

\title{\textbf{On Reconstructing Conservative and Primitive Variables: An Eigenvector Analysis on Curvilinear Grids}}
\author{Amareshwara Sainadh Chamarthi\\
\small Division of Engineering and Applied Science\\
\small California Institute of Technology, Pasadena, CA 91125, USA}

\date{April 2026}

\begin{document}
\maketitle

\begin{abstract}
In wall-modelled large-eddy simulations of hypersonic boundary-layer transition, Hoffmann, Chamarthi and Frankel reported that characteristic reconstruction based on conservative-variable eigenvectors produced markedly better results than the corresponding primitive-variable implementation. The observation was empirical. A subsequent wave-appropriate conservative reconstruction (WA-CR) algorithm used a rank-one entropy correction based on the premise that contact-discontinuity error lies in a single conservative entropy/contact direction. This note gives the algebraic foundation for both observations. For the standard conservative curvilinear eigenvectors, the density row of the right-eigenvector matrix contains exact, metric-free zeros in the shear columns, so shear waves carry no density perturbation and a contact discontinuity is represented by the conservative entropy eigenvector alone. The conservative left eigenvectors provide the dual projection property: the entropy amplitude is obtained with a metric-independent left eigenvector and has unit contact scaling, while total-energy perturbations have zero projection onto the shear amplitudes. In the standard primitive curvilinear eigenvectors, by contrast, shear right eigenvectors contain metric-dependent density components and the primitive entropy left eigenvector contains metric-weighted tangential-velocity terms. Thus the conservative formulation supplies the two algebraic requirements for an exact, sufficient, metric-invariant, rank-one entropy correction: metric-independent entropy projection and a metric-independent entropy update direction. Curvilinear metrics make the distinction explicit, but the conservative state-space contact direction is already the natural direction underlying WA-CR even on Cartesian grids.
\end{abstract}

\section{Introduction}

Characteristic-space reconstruction is a standard mechanism by which high-resolution finite-volume and finite-difference schemes reduce oscillations in compressible Euler and Navier--Stokes calculations~\cite{van2006upwind}. The transformation from physical variables to characteristic variables requires left and right eigenvectors of the convective flux Jacobian. This transformation can be performed using conservative variables, denoted by $\mathbf{U}$, or primitive variables, denoted by $\mathbf{W}$. Since the conservative and primitive Jacobians (\textit{it's not appropriate to call them Jacobians for primitive variables; it's an abuse of terminology}) share similarities, their eigenvalues are the same.  It is therefore tempting to regard the two characteristic projections as essentially interchangeable.

In the sequence of papers on wave-appropriate reconstruction~\cite{chamarthi2023wave,Hoffmann2024,Chamarthi2025Multiphase,Chamarthi2025Interface,Chamarthi2026Rank1}, the apparent interchangeability of these methods became questionable. In Ref.~\cite{Hoffmann2024}, the same centralized gradient-based reconstruction was implemented in two forms: MEG-C-CONS, which constructs characteristic variables from conservative variables, and MEG-C-PRIM, which constructs them from primitive variables (MEG-Monotonicity preserving explicit gradient-based reconstruction). In Mach~6 oblique-shock/boundary-layer interaction and Mach~7.7 compression-ramp simulations on clustered body-fitted grids, the conservative-variable version was found to be crucial for accurately predicting the laminar-to-turbulent transition. While the paper reported this observation, it did not provide an algebraic explanation. The later WA-CR (Wave-appropriate conservative variable reconstruction) algorithm~\cite{Chamarthi2026Rank1} further complicated the issue. WA-CR corrects contact-discontinuity errors by a rank-one update along the entropy/contact characteristic. This relies on the assumption that a contact discontinuity is carried by a single characteristic wave, not a mixture of entropy and shear waves. In Cartesian coordinates, this may be straightforward for the conservative Euler eigenvectors. However, on curvilinear grids, metric coefficients are incorporated into the eigenvector matrices, and it is not immediately clear whether the same rank-one structure remains intact.

This note illustrates that the concept persists in the conservative formulation, but not in the standard primitive eigenbasis. The reason behind this is the zero structure of the conservative right and left eigenvectors. On a curvilinear grid, the conservative density row contains exact zeros in the shear columns, which means that shear waves do not cause any density perturbation. The conservative entropy right eigenvector represents the physical contact direction.
\begin{equation}
  \mathbf{r}_2^{\mathrm{cons}} = \left(1,u,v,w,\frac{1}{2}q^2\right)^T,
\end{equation}
and contains no metric coefficients. The corresponding conservative entropy left eigenvector projects the contact amplitude without metric-weighted tangential-velocity coupling. These two properties are precisely the projection and update requirements of a rank-one correction.

The standard primitive eigenvectors used in curvilinear compressible-flow codes have a different structure. Their density row contains metric-dependent shear entries, and their entropy left eigenvector contains metric-weighted tangential velocity terms. Thus a contact discontinuity is not represented, in that standard primitive basis, by a metric-invariant entropy direction alone. This explains the empirical conservative-versus-primitive disparity of Ref.~\cite{Hoffmann2024} and supplies the curvilinear algebraic foundation for WA-CR.

Curvilinear grids expose the distinction most visibly because metric coefficients enter the standard primitive eigenvectors, but the conservative construction is not merely a curvilinear fix. Even on Cartesian grids, the contact jump is naturally represented in conservative state space as $\Delta\rho(1,u,v,w,\tfrac{1}{2}q^2)^T$, and WA-CR is formulated around projecting and updating precisely that conservative entropy/contact direction.

\section{Curvilinear Euler Equations and Variable Sets}\label{sec:setup}

The compressible Euler equations are written on curvilinear coordinates $(\xi,\eta,\zeta)$ as
\begin{equation}
  \frac{\partial \mathbf{U}}{\partial t}
  + \frac{\partial \bar{\mathbf{F}}}{\partial \xi}
  + \frac{\partial \bar{\mathbf{G}}}{\partial \eta}
  + \frac{\partial \bar{\mathbf{H}}}{\partial \zeta} = 0.
  \label{eq:governing}
\end{equation}
For the present algebraic discussion the scalar Jacobian factor does not alter the eigenvector structure, so we write the physical conservative and primitive state vectors as
\begin{equation}
  \mathbf{U}=\left(\rho,\rho u,\rho v,\rho w,\rho E\right)^T,
  \qquad
  \mathbf{W}=\left(\rho,u,v,w,p\right)^T.
\end{equation}
Following Haselbacher and Landmann~\cite{Haselbacher}, the quasilinear Jacobian in the $\xi$ direction is formed with the unnormalised metrics $\xi_x,\xi_y,\xi_z$. The corresponding contravariant velocity is
\begin{equation}
  \bar{u}=\xi_x u+\xi_y v+\xi_z w.
\end{equation}
Let
\begin{equation}
  \beta_\xi = \|\nabla\xi\|=\sqrt{\xi_x^2+\xi_y^2+\xi_z^2},
  \qquad
  \hat{\bm{\xi}}=\frac{\nabla\xi}{\|\nabla\xi\|},
  \qquad
  \hat{u}=\hat{\xi}_x u+\hat{\xi}_y v+\hat{\xi}_z w.
\end{equation}
We denote the thermodynamic sound speed by $c$ and the metric-scaled acoustic speed by
\begin{equation}
  a_\xi = c\,\beta_\xi .
\end{equation}
This distinction is important: $a_\xi$ appears in the eigenvalues, whereas the right and left eigenvector normalisations below use the physical sound speed $c$. The five eigenvalues of the $\xi$-direction flux Jacobian are
\begin{equation}
  \lambda_1=\bar{u}-a_\xi,\qquad
  \lambda_{2,3,4}=\bar{u},\qquad
  \lambda_5=\bar{u}+a_\xi.
  \label{eq:eigenvalues}
\end{equation}
The triple eigenvalue corresponds to the entropy wave and two shear waves.
The identities used below are identities for characteristic reconstruction of the state vector. They should not be interpreted as identities for reconstructed physical flux components: for a perturbation $\Delta\mathbf{U}=\mathbf{R}\Delta\mathbf{C}$, the corresponding flux perturbation is $\Delta\mathbf{F}=\mathbf{A}\Delta\mathbf{U}=\mathbf{R}\mathbf{\Lambda}\Delta\mathbf{C}$ and therefore carries eigenvalue-weighted characteristic amplitudes.

\section{The Similarity Transform Carries no Grid Metrics}\label{sec:similarity}

Let $\mathbf{M}=\partial\mathbf{U}/\partial\mathbf{W}$ be the primitive-to-conservative transformation Jacobian. A direct computation gives
\begin{equation}
  \mathbf{M}=
  \begin{pmatrix}
    1 & 0 & 0 & 0 & 0 \\
    u & \rho & 0 & 0 & 0 \\
    v & 0 & \rho & 0 & 0 \\
    w & 0 & 0 & \rho & 0 \\
    \tfrac{1}{2}q^2 & \rho u & \rho v & \rho w & \tfrac{1}{\gamma-1}
  \end{pmatrix},
  \qquad q^2=u^2+v^2+w^2.
  \label{eq:M}
\end{equation}
The primitive and conservative Jacobians are similar,
\begin{equation}
  \bar{\mathbf{A}}_p=\mathbf{M}^{-1}\bar{\mathbf{A}}\mathbf{M},
\end{equation}
so they have the same eigenvalues. However, $\mathbf{M}$ and $\mathbf{M}^{-1}$ contain only local thermodynamic and velocity variables. They contain no grid metrics $\hat{\xi}_x,\hat{\xi}_y,\hat{\xi}_z$. Consequently, metric dependence found in one of the standard eigenvector matrices is not created or removed by this state-variable similarity transform alone. It is a feature of the chosen eigenbasis, especially inside the degenerate eigenspace associated with $\lambda=\bar{u}$.

\begin{keybox}[What must be compared]
The eigenvalues are identical; the eigenvectors are not. For wave-appropriate reconstruction, the question is not which variable set has the correct wave speeds, but whether the eigenvectors give a metric-independent decoupling between entropy/contact and shear content.
\end{keybox}

\section{Right Eigenvectors: State variable update}\label{sec:right}

The right eigenvector $\mathbf{r}_b$ represents the state perturbation corresponding to a specific characteristic amplitude. If the characteristic amplitude $C_b$ undergoes a change of $\delta$, which physical variables are affected or undergo a change?

\subsection{Conservative variables}

Using the standard Haselbacher--Landmann ordering~\cite{Haselbacher} (acoustic$^-$, entropy, shear$_1$, shear$_2$, acoustic$^+$), the conservative right-eigenvector matrix in the $\xi$ direction is
\begin{equation}
\bar{\mathbf{R}}_\xi^{\mathrm{cons}}=
\begin{pmatrix}
1 & 1 & 0 & 0 & 1 \\
 u-c\hat{\xi}_x & u & -\hat{\xi}_y & -\hat{\xi}_z & u+c\hat{\xi}_x \\
 v-c\hat{\xi}_y & v & \hat{\xi}_x & 0 & v+c\hat{\xi}_y \\
 w-c\hat{\xi}_z & w & 0 & \hat{\xi}_x & w+c\hat{\xi}_z \\
 H-c\hat{u} & \tfrac{1}{2}q^2 & v\hat{\xi}_x-u\hat{\xi}_y & w\hat{\xi}_x-u\hat{\xi}_z & H+c\hat{u}
\end{pmatrix}.
\label{eq:Rcons}
\end{equation}
The density row is
\begin{equation}
  \boxed{(1,\;1,\;0,\;0,\;1)}.
  \label{eq:Rcons_density_row}
\end{equation}
The two shear columns have exact zeros in the density row. These zeros are independent of the metric coefficients.

\subsection{Primitive variables}

The standard primitive right-eigenvector matrix in the same reference~\cite{Haselbacher} is
\begin{equation}
\bar{\mathbf{R}}_\xi^{\mathrm{prim}}=
\begin{pmatrix}
\dfrac{1}{2c^2} & \dfrac{\hat{\xi}_x}{c^2} & \dfrac{\hat{\xi}_z}{c^2} & -\dfrac{\hat{\xi}_y}{c^2} & \dfrac{1}{2c^2} \\[10pt]
-\dfrac{\hat{\xi}_x}{2\rho c} & 0 & -\hat{\xi}_y & -\hat{\xi}_z & \dfrac{\hat{\xi}_x}{2\rho c} \\[8pt]
-\dfrac{\hat{\xi}_y}{2\rho c} & -\hat{\xi}_z & \hat{\xi}_x & 0 & \dfrac{\hat{\xi}_y}{2\rho c} \\[8pt]
-\dfrac{\hat{\xi}_z}{2\rho c} & \hat{\xi}_y & 0 & \hat{\xi}_x & \dfrac{\hat{\xi}_z}{2\rho c} \\[8pt]
\dfrac{1}{2} & 0 & 0 & 0 & \dfrac{1}{2}
\end{pmatrix}.
\label{eq:Rprim}
\end{equation}
The density row is
\begin{equation}
  \boxed{\left(\frac{1}{2c^2},\;\frac{\hat{\xi}_x}{c^2},\;\frac{\hat{\xi}_z}{c^2},\;-\frac{\hat{\xi}_y}{c^2},\;\frac{1}{2c^2}\right)}.
  \label{eq:Rprim_density_row}
\end{equation}
The shear columns now carry density content proportional to $\hat{\xi}_z/c^2$ and $-\hat{\xi}_y/c^2$. On non-axis-aligned faces these entries are generally non-zero.

\begin{resultbox}[Right-eigenvector result]
In the conservative basis, shear waves are density-neutral on every curvilinear grid. In the standard primitive basis, shear waves carry metric-dependent density content. Thus conservative variables provide metric-independent entropy-shear decoupling on the update side, while the standard primitive basis does not.
\end{resultbox}

\section{The Contact Discontinuity Test}\label{sec:contact}

A contact discontinuity satisfies
\begin{equation}
  [\rho]\neq 0,
  \qquad [u]=0,
  \qquad [p]=0.
  \label{eq:contact_definition}
\end{equation}
It is a pure density jump at constant velocity and pressure. In viscous flows, the tangential velocities are continuous, as mentioned in \cite{batchelor1967introduction}, or in the presence of artificial viscosity, as discussed in \cite{meng2018numerical}. Additionally, the analysis reveals that shear waves do not affect the density when conservative-characteristic variables are reconstructed.

\subsection{Conservative variables}

The jump in conservative variables is
\begin{equation}
  \Delta\mathbf{U}
  =\Delta\rho\left(1,u,v,w,\frac{1}{2}q^2\right)^T
  =\Delta\rho\,\mathbf{r}_2^{\mathrm{cons}}.
  \label{eq:contact_cons}
\end{equation}
Thus the contact maps onto exactly one characteristic: the conservative entropy eigenvector. No metric coefficient appears. The decomposition is rank-one at every face of every curvilinear grid.

\subsection{Primitive variables with the standard basis}

The same contact jump in primitive variables is
\begin{equation}
  \Delta\mathbf{W}=(\Delta\rho,0,0,0,0)^T.
\end{equation}
To represent it using the standard primitive basis one solves
\begin{equation}
  \bar{\mathbf{R}}_\xi^{\mathrm{prim}}\mathbf{c}=\Delta\mathbf{W}.
\end{equation}
Equivalently, projecting the primitive contact jump with Eq.~\eqref{eq:Rinvprim} gives
\begin{equation}
  \Delta\mathbf{C}^{\mathrm{prim}}
  = (\bar{\mathbf{R}}_\xi^{\mathrm{prim}})^{-1}\Delta\mathbf{W}
  = c^2\Delta\rho\,
  \left(0,\;\hat{\xi}_x,\;\hat{\xi}_z,\;-\hat{\xi}_y,\;0\right)^T.
  \label{eq:primitive_contact_projection}
\end{equation}
Thus the contact jump is distributed over the standard primitive entropy and shear amplitudes. The factor $c^2$ is a normalization factor associated with the Haselbacher--Landmann primitive eigenvectors; the essential point is the metric-dependent distribution over $\hat{\xi}_x$, $\hat{\xi}_z$, and $-\hat{\xi}_y$. Substitution into the density row of Eq.~\eqref{eq:Rprim} verifies the reconstruction,
\begin{equation}
  \frac{\hat{\xi}_x}{c^2}\left(c^2\hat{\xi}_x\Delta\rho\right)
  +\frac{\hat{\xi}_z}{c^2}\left(c^2\hat{\xi}_z\Delta\rho\right)
  -\frac{\hat{\xi}_y}{c^2}\left(-c^2\hat{\xi}_y\Delta\rho\right)
  =\left(\hat{\xi}_x^2+\hat{\xi}_y^2+\hat{\xi}_z^2\right)\Delta\rho
  =\Delta\rho.
  \label{eq:primitive_contact_check}
\end{equation}
Therefore the primitive contact is represented by a face-orientation-dependent combination inside the repeated eigenvalue subspace, not by the standard primitive entropy column alone.

\begin{warnbox}[Contact is not rank-one in the standard primitive basis]
In conservative variables, a contact discontinuity is exactly $\Delta\rho\,\mathbf{r}_2^{\mathrm{cons}}$. In the standard primitive curvilinear eigenbasis, the same physical jump is represented by a metric-dependent superposition inside the degenerate entropy-shear eigenspace. A one-direction entropy wave correction is therefore not naturally available in this standard primitive form.
\end{warnbox}

\section{Left Eigenvectors: Characteristic Projection}
\label{sec:left}

The analysis so far has concentrated on the \emph{right} eigenvectors, which are the columns of $\bar{\mathbf{R}}_\xi$ that describe how each characteristic wave \emph{influences} the physical state. The \emph{left} eigenvectors, on the other hand, represent the rows of $\bar{\mathbf{R}}^{-1}_\xi$ and address the complementary question:
\begin{quote}
  \textit{Given the physical state, how is the amplitude of
  each characteristic wave obtained by projection?}
\end{quote}
The characteristic amplitude of wave $b$ is obtained by projecting the state onto the
$b$-th left eigenvector:
\begin{equation}
  C_b = \mathbf{l}_b \cdot \mathbf{U}
  \quad\text{(conservative)}
  \qquad\text{or}\qquad
  C_b = \mathbf{l}_b \cdot \mathbf{W}
  \quad\text{(primitive).}
\end{equation}
In the WA-CR algorithm, both the stencil amplitudes
$C_{2,j} = \mathbf{l}_2\cdot\mathbf{U}_{i+j}$
(Eq.~\eqref{eq:wacr_stencil}) and the deficit
$\delta^L = \hat{C}^L_2 - \mathbf{l}_2\cdot\mathbf{U}^{L,c}_{i+\nicefrac{1}{2}}$
(Eq.~\eqref{eq:wacr_deficit}) depend critically on the properties of $\mathbf{l}_2$.
Metric-dependent coupling in $\mathbf{l}_2$ propagates to every projected stencil amplitude and to the amplitude defect in Eq.~\eqref{eq:wacr_deficit}.  The right and left eigenvectors are therefore equally important.

\bigskip
\subsection*{Left Eigenvectors for Conservative Variables}

The full inverse $(\bar{\mathbf{R}}^{\,\mathrm{cons}}_\xi)^{-1}$, written with the
overall prefactor $(\gamma{-}1)/(2c^2)$ factored out, is
(where $\hat\gamma = \gamma-1$ and $\hat{u} = u\hat\xi_x + v\hat\xi_y + w\hat\xi_z$):

\begin{equation}
  (\bar{\mathbf{R}}^{\,\mathrm{cons}}_\xi)^{-1}
  = \frac{\gamma-1}{2c^2}
  \begin{pmatrix}
    \dfrac{c\hat{u}}{\hat\gamma}+\dfrac{q^2}{2}
      & {-}\dfrac{c\hat\xi_x}{\hat\gamma}{-}u
      & {-}\dfrac{c\hat\xi_y}{\hat\gamma}{-}v
      & {-}\dfrac{c\hat\xi_z}{\hat\gamma}{-}w & 1 \\[8pt]
    \dfrac{4c^2}{\hat\gamma}-2H & 2u & 2v & 2w & -2 \\[8pt]
    \dfrac{2c^2(\hat\xi_y\hat{u}-v)}{\hat\gamma\hat\xi_x}
      & {-}\dfrac{2c^2\hat\xi_y}{\hat\gamma}
      & \dfrac{2c^2(\hat\xi_x^2+\hat\xi_z^2)}{\hat\gamma\hat\xi_x}
      & {-}\dfrac{2c^2\hat\xi_z\hat\xi_y}{\hat\gamma\hat\xi_x}
      & \mathbf{0} \\[8pt]
    \dfrac{2c^2(\hat\xi_z\hat{u}-w)}{\hat\gamma\hat\xi_x}
      & {-}\dfrac{2c^2\hat\xi_z}{\hat\gamma}
      & {-}\dfrac{2c^2\hat\xi_z\hat\xi_y}{\hat\gamma\hat\xi_x}
      & \dfrac{2c^2(\hat\xi_x^2+\hat\xi_y^2)}{\hat\gamma\hat\xi_x}
      & \mathbf{0} \\[8pt]
    {-}\dfrac{c\hat{u}}{\hat\gamma}+\dfrac{q^2}{2}
      & \dfrac{c\hat\xi_x}{\hat\gamma}{-}u
      & \dfrac{c\hat\xi_y}{\hat\gamma}{-}v
      & \dfrac{c\hat\xi_z}{\hat\gamma}{-}w & 1
  \end{pmatrix}
  \label{eq:Rinvcons}
\end{equation}

\noindent\textbf{Focus on Row~2 (entropy left eigenvector):}
\begin{equation}
  \boxed{
    \mathbf{l}_2^{\,\mathrm{cons}}
    = \frac{\gamma-1}{2c^2}
    \Bigl(\tfrac{4c^2}{\hat\gamma}-2H,\quad 2u,\quad 2v,\quad 2w,\quad -2\Bigr)
  }
  \label{eq:l2cons}
\end{equation}
\textbf{No metric coefficient $\hat\xi_x,\hat\xi_y,\hat\xi_z$ appears.}
The entropy amplitude at any stencil point is
\begin{equation}
  C_2 = \mathbf{l}_2^{\,\mathrm{cons}}\cdot\mathbf{U}
  = \frac{\gamma-1}{2c^2}
  \Bigl[\Bigl(\tfrac{4c^2}{\hat\gamma}-2H\Bigr)\rho
  + 2(\rho u)\,u + 2(\rho v)\,v + 2(\rho w)\,w - 2\,\rho E\Bigr].
\end{equation}
For a contact discontinuity, we showed in Section~\ref{sec:contact} that
$\Delta\mathbf{U} = \Delta\rho\cdot\mathbf{r}_2^{\mathrm{cons}}$.
By the biorthogonality condition $\mathbf{l}_b\cdot\mathbf{r}_c = \delta_{bc}$,
\begin{equation}
  \Delta C_2^{\mathrm{cons}} = \mathbf{l}_2^{\mathrm{cons}}\cdot\Delta\mathbf{U}
  = \Delta\rho\;\mathbf{l}_2^{\mathrm{cons}}\cdot\mathbf{r}_2^{\mathrm{cons}}
  = \Delta\rho.
  \label{eq:contact_projection}
\end{equation}
The projection is exact, carries a unit scale factor, and is independent of the
grid orientation.

\noindent\textbf{Focus on the last column of Eq.~\eqref{eq:Rinvcons} (acting on
$\rho E$):}
\begin{equation}
  \text{Last column of }(\bar{\mathbf{R}}^{\,\mathrm{cons}}_\xi)^{-1}
  \propto
  \bigl(1,\;\; -2,\;\; \mathbf{0},\;\; \mathbf{0},\;\; 1\bigr)^T.
  \label{eq:rhoE_col}
\end{equation}
Entries 3 and 4 (shear rows) are \textbf{exact zeros}: a change in total energy
$\rho E$ does not alter the shear characteristic amplitudes $C_3$ or $C_4$.
This is the \emph{column-dual} of the density-row zeros in $\bar{\mathbf{R}}^{\,\mathrm{cons}}_\xi$:

\begin{center}
\renewcommand{\arraystretch}{1.5}
\begin{tabular}{lll}
\toprule
Direction & Statement & Location \\
\midrule
Right (columns of $\mathbf{R}$)
  & Shear waves do not change $\rho$
  & Zeros in Row~1 of $\bar{\mathbf{R}}^{\,\mathrm{cons}}_\xi$ \\
Left (rows of $\mathbf{R}^{-1}$)
  & $\rho E$ changes do not project onto shear modes
  & Zeros in last column of $(\bar{\mathbf{R}}^{\,\mathrm{cons}}_\xi)^{-1}$ \\
\bottomrule
\end{tabular}
\end{center}

\bigskip
\subsection*{Left Eigenvectors for Primitive Variables}

\begin{equation}
  (\bar{\mathbf{R}}^{\,\mathrm{prim}}_\xi)^{-1}
  =
  \begin{pmatrix}
    0 & -\hat\xi_x\rho c & -\hat\xi_y\rho c & -\hat\xi_z\rho c & 1 \\[4pt]
    \hat\xi_x c^2 & 0 & -\hat\xi_z & \hat\xi_y & -\hat\xi_x \\[4pt]
    \hat\xi_z c^2 & -\hat\xi_y & \hat\xi_x & 0 & -\hat\xi_z \\[4pt]
    -\hat\xi_y c^2 & -\hat\xi_z & 0 & \hat\xi_x & \hat\xi_y \\[4pt]
    0 & \hat\xi_x\rho c & \hat\xi_y\rho c & \hat\xi_z\rho c & 1
  \end{pmatrix}
  \label{eq:Rinvprim}
\end{equation}

\noindent\textbf{Focus on Row~2 (entropy left eigenvector):}
\begin{equation}
  \boxed{
    \mathbf{l}_2^{\,\mathrm{prim}}
    = \Bigl(\hat\xi_x c^2,\quad 0,\quad -\hat\xi_z,\quad \hat\xi_y,\quad -\hat\xi_x\Bigr)
  }
  \label{eq:l2prim}
\end{equation}
\textbf{Metric coefficients appear in every non-trivially-zero entry.}
The entropy amplitude is
\begin{equation}
  C_2^{\mathrm{prim}}
  = \hat\xi_x c^2\rho - \hat\xi_z v + \hat\xi_y w - \hat\xi_x p
  = \hat\xi_x(c^2\rho - p) - \hat\xi_z v + \hat\xi_y w.
  \label{eq:C2prim}
\end{equation}
On a Cartesian grid ($\hat\xi_x=1,\,\hat\xi_y=\hat\xi_z=0$) this reduces correctly
to $C_2^{\mathrm{prim}} = c^2\rho - p$, the standard primitive entropy proxy.
On a curved grid, however, the projection additionally picks up the tangential velocity
components $v$ and $w$ with metric-dependent weights $\hat\xi_y$ and $-\hat\xi_z$ ---
and these weights differ at every grid face.

\noindent\textbf{Focus on the last column of Eq.~\eqref{eq:Rinvprim} (acting on $p$):}
\begin{equation}
  \text{Last column of }(\bar{\mathbf{R}}^{\,\mathrm{prim}}_\xi)^{-1}
  =
  \bigl(1,\;\; -\hat\xi_x,\;\; \mathbf{-\hat\xi_z},\;\; \mathbf{+\hat\xi_y},\;\; 1\bigr)^T.
  \label{eq:p_col}
\end{equation}
Entries 3 and 4 (shear rows) are $-\hat\xi_z$ and $+\hat\xi_y$, which are generally nonzero on non-axis-aligned curvilinear faces.  A pressure perturbation therefore has nonzero projection onto the shear
characteristic amplitudes:
\begin{equation}
  \frac{\partial C_3^{\mathrm{prim}}}{\partial p} = -\hat\xi_z \neq 0,
  \qquad
  \frac{\partial C_4^{\mathrm{prim}}}{\partial p} = +\hat\xi_y \neq 0.
\end{equation}
This mirrors the metric coupling already found in $\bar{\mathbf{R}}^{\,\mathrm{prim}}_\xi$:
just as shear right eigenvectors carry density (metric coupling in the
\emph{forward} direction), pressure changes produce nonzero shear characteristic amplitudes
(metric coupling in the \emph{inverse} direction).

\begin{warnbox}[Primitive entropy projection contains metric-weighted velocity terms]
In primitive variables on a curved grid, the entropy amplitude
$C_2^{\mathrm{prim}}$ contains the tangential velocity contributions
$-\hat\xi_z v + \hat\xi_y w$ (Eq.~\eqref{eq:C2prim}).
For a genuine contact ($[v]=[w]=0$) these terms happen not to jump, so
Eq.~\eqref{eq:primitive_contact_projection} gives a nonzero primitive entropy component, but with the standard-normalization amplitude $c^2\hat\xi_x\Delta\rho$ and with simultaneous shear components $c^2\hat\xi_z\Delta\rho$ and $-c^2\hat\xi_y\Delta\rho$ that vary from face to face.
More critically, in any stencil with metric-weighted tangential-velocity variation, the stencil amplitudes $C_{2,j}^{\mathrm{prim}}$ mix entropy content with shear
information that has no bearing on the contact discontinuity being corrected.
The WA-CR stencil amplitude is no longer a pure entropy/contact measure even when no entropy variation is present.
\end{warnbox}

The comparison between the two entropy left eigenvectors, Eqs.~\eqref{eq:l2cons}
and~\eqref{eq:l2prim}, is summarised below.

\begin{center}
\renewcommand{\arraystretch}{1.6}
\begin{tabular}{p{3.5cm}p{4.5cm}p{4.5cm}}
\toprule
\textbf{Property of $\mathbf{l}_2$}
  & \textbf{Conservative}
  & \textbf{Primitive (curved grid)} \\
\midrule
Metric dependence
  & None (Eq.~\eqref{eq:l2cons})
  & $\hat\xi_x, \hat\xi_y, \hat\xi_z$ in every entry (Eq.~\eqref{eq:l2prim}) \\
Contact amplitude $\Delta C_2$
  & $= \Delta\rho$ (exact, unit-scaled)
  & $= c^2\hat\xi_x\,\Delta\rho$ in the standard normalization, with simultaneous shear components $c^2\hat\xi_z\Delta\rho$ and $-c^2\hat\xi_y\Delta\rho$ \\
Velocity coupling
  & No metric-weighted tangential velocity coupling
  & $-\hat\xi_z v + \hat\xi_y w$ at every stencil point \\
Pressure $\to$ shear coupling
  & Zero (Eq.~\eqref{eq:rhoE_col})
  & $-\hat\xi_z,\,+\hat\xi_y \neq 0$ (Eq.~\eqref{eq:p_col}) \\
Stencil quality
  & Measures pure entropy content
  & Mixes entropy with shear kinematics \\
\bottomrule
\end{tabular}
\end{center}


\section{Zeros in eigenvectors}\label{sec:duality}

Every valid eigenbasis satisfies biorthogonality. What distinguishes the conservative basis used here is that the zeros needed for WA-CR appear explicitly, simultaneously, and without metric coefficients.
\begin{equation}
  \mathbf{l}_b\cdot\mathbf{r}_c=\delta_{bc},
  \qquad
  \bar{\mathbf{R}}_\xi^{-1}\bar{\mathbf{R}}_\xi=\mathbf{I}.
\end{equation}
\begin{center}
\renewcommand{\arraystretch}{1.45}
\begin{tabular}{p{3.5cm}p{5.1cm}p{5.1cm}}
\toprule
 & \textbf{Conservative} & \textbf{Standard primitive on curved grid} \\
\midrule
Right eigenvectors & Shear columns have zero density row; shear waves do not change $\rho$. & Shear columns have density entries $\hat{\xi}_z/c^2$ and $-\hat{\xi}_y/c^2$. \\
Left eigenvectors & Energy column has zero shear rows; energy changes have zero projection onto shear amplitudes. & Pressure column has shear entries $-\hat{\xi}_z$ and $+\hat{\xi}_y$. \\
Entropy projection & Metric-free $\mathbf{l}_2^{\mathrm{cons}}$; contact amplitude is $\Delta\rho$. & Metric-dependent $\mathbf{l}_2^{\mathrm{prim}}$; entropy amplitude contains tangential velocity terms. \\
Entropy update & Metric-free $\mathbf{r}_2^{\mathrm{cons}}$; correction is a physical contact jump. & Standard primitive update is metric-dependent and mixed with shear. \\
\bottomrule
\end{tabular}
\end{center}

\begin{keybox}[Projection and update requirements]
A rank-one entropy correction necessitates two independent algebraic properties. Firstly, the projected amplitude $C_2=\mathbf{l}_2\cdot\mathbf{U}$ must accurately represent entropy or contact content at each stencil point, excluding any metric-weighted shear contributions. Secondly, the update $\delta\mathbf{r}_2$ must modify the face state along the entropy or contact direction without altering the shear or acoustic characteristics. Conservative variables inherently satisfy both requirements within the standard curvilinear eigenbasis. However, the standard primitive basis does not.
\end{keybox}

\section{Consequence for WA-CR}\label{sec:wacr}

The WA-CR update can be written schematically as
\begin{align}
  C_{2,j} &= \mathbf{l}_2^{\mathrm{cons}}\cdot\mathbf{U}_{i+j},
  \qquad j\in\{-2,\ldots,3\},\label{eq:wacr_stencil}\\
  \delta^L &= \widehat{C}_{2}^{L}-\mathbf{l}_2^{\mathrm{cons}}\cdot\mathbf{U}_{i+1/2}^{L,c},\label{eq:wacr_deficit}\\
  \mathbf{U}_{i+1/2}^{L}
    &=\mathbf{U}_{i+1/2}^{L,c}+\delta^L\mathbf{r}_2^{\mathrm{cons}}.
  \label{eq:wacr_update}
\end{align}
Here $\mathbf{U}_{i+1/2}^{L,c}$ is a candidate conservative interface state and $\widehat{C}_{2}^{L}$ is the reconstructed entropy amplitude. The update is exact and sufficient because
\begin{equation}
  \mathbf{r}_2^{\mathrm{cons}}=\left(1,u,v,w,\frac{1}{2}q^2\right)^T
\end{equation}
contains no grid metric and is the conservative representation of a contact jump. It is also minimal cost: once $\delta^L$ is known, the correction is one scalar times one vector.

The analysis above shows that the curvilinear extension is algorithmically unchanged. The same entropy projection and the same update vector apply at every grid face. No face-dependent distribution of shear amplitudes is required. For the standard primitive basis, the corresponding correction would not have this form. The projected entropy amplitude contains metric-weighted tangential velocity terms, and the right-eigenvector representation of a contact is not entropy-only. A correction would require a face-dependent combination of entropy and shear directions. That is not WA-CR in its rank-one form.

\section{Conclusions}\label{sec:conclusions}

This note gives an algebraic explanation for the observed superiority of conservative-variable characteristic reconstruction in the wave-appropriate reconstruction program and for the rank-one structure of WA-CR. The key point is not the eigenvalues: conservative and primitive Jacobians have the same wave speeds. The key point is the eigenvector zero structure.

The conservative right eigenvectors have exact, metric-free zeros in the density row of the shear columns. Therefore shear waves do not carry density perturbations, and a contact discontinuity is exactly
\begin{equation}
  \Delta\mathbf{U}=\Delta\rho\left(1,u,v,w,\frac{1}{2}q^2\right)^T.
\end{equation}
The conservative left eigenvectors provide the dual property: entropy projection is metric-free, gives $\Delta C_2=\Delta\rho$ for a contact, and does not mix shear amplitudes through the energy column. Thus both operations required by WA-CR, projection with $\mathbf{l}_2$ and update with $\mathbf{r}_2$, are metric-independent, decoupled, and rank-one. The standard primitive curvilinear eigenbasis lacks this paired zero structure. Its shear right eigenvectors carry density entries proportional to grid metrics, its entropy left eigenvector contains metric-weighted tangential velocity terms, and its pressure column has nonzero projection onto shear amplitudes. A contact discontinuity is therefore not represented as a single metric-free entropy direction in that standard primitive basis. The correction would require face-dependent redistribution into entropy and shear amplitudes, which is not the WA-CR rank-one update.

The conclusions of this note are likely independent of the specific reconstruction scheme employed after characteristic projection. They solely stem from the eigenstructure of the compressible Euler flux Jacobian. Regardless of whether the interface states are reconstructed using any high-order method, the characteristic variables are obtained by projecting state perturbations onto a predetermined eigenbasis. Consequently, the physical interpretation of the reconstructed quantities is determined prior to the application of any limiter or polynomial reconstruction. In the conservative eigensystem, the acoustic, entropy/contact, and shear/vortical components correspond to decoupled conservative-state perturbations with a metric-independent contact direction and density-neutral shear directions. However, in the standard primitive eigensystem on curvilinear grids, this separation is lost due to metric-dependent coupling between the entropy and shear components.

\subsection*{Summary Table}
\begin{center}
\renewcommand{\arraystretch}{1.45}
\begin{tabular}{p{4.2cm}p{4.7cm}p{4.7cm}}
\toprule
\textbf{Property} & \textbf{Conservative} & \textbf{Standard primitive on curved grid} \\
\midrule
Shear $\to$ density, right eigenvectors & Zero, exact and metric-free & $\hat{\xi}_z/c^2$, $-\hat{\xi}_y/c^2$ \\
Energy/pressure $\to$ shear amplitude, left eigenvectors & Zero, exact and metric-free & $-\hat{\xi}_z$, $+\hat{\xi}_y$ \\
Contact jump $\to$ characteristic basis & Rank-one, entropy only & Entropy-shear superposition \\
Entropy right eigenvector & $(1,u,v,w,\tfrac{1}{2}q^2)^T$, no metrics & Metric-dependent in standard basis \\
Entropy left eigenvector & Metric-free; projects $\Delta C_2=\Delta\rho$ & Metric-dependent; contains tangential velocity terms \\
WA-CR rank-one correction & Exact, sufficient, metric-invariant & Not available in this standard form \\
Grid generality & Every curvilinear grid & Requires rebasing or face-dependent correction \\
\bottomrule
\end{tabular}
\end{center}

\section*{Acknowledgements}\label{acknow}
This note was written as supplementary material for the WA-CR algorithm presented in Ref.~\cite{Chamarthi2026Rank1}. The author gratefully acknowledges many discussions with Natan Hoffmann, Prof. Steven H. Frankel, and Sean Bokor over the years, which helped motivate and clarify the conservative-variable characteristic reconstruction perspective explained here. The author is especially grateful to Natan Hoffmann and Sean Bokor for their patience and substantial implementation efforts during the development and testing of these reconstruction algorithms presented and published over the years. The author understands that this process may, at times, have been frustrating. To borrow from football and Sir Alex Ferguson, science too can sometimes be painfully difficult.


\end{document}